\documentclass[11pt,oneside,reqno]{amsart}
\usepackage[T1]{fontenc}
\usepackage[latin9]{inputenc}
\usepackage[a4paper]{geometry}
\usepackage{color}
\usepackage{float}
\usepackage{amstext}
\usepackage{amssymb}
\usepackage{graphicx}
\usepackage{setspace}
\makeatletter

\providecommand{\tabularnewline}{\\}

\usepackage{amsfonts}
\usepackage{amstext}

\newtheorem{thm}{Theorem}

\newtheorem{prop}{Proposition}
\newtheorem*{asm}{Assumption}

\theoremstyle{definition}
\newtheorem{rem}{Remark}

\makeatother

\usepackage[style=authoryear,backend=bibtex]{biblatex}
\begin{document}
\title{Conditional likelihood ratio test with many weak instruments}
\author{Sreevidya Ayyar}
\address{Department of Economics, London School of Economics, Houghton Street,
London, WC2A 2AE, UK.}
\email{s.ayyar1@lse.ac.uk}
\author{Yukitoshi Matsushita}
\address{Graduate School of Economics, Hitotsubashi University, 2-1 Naka, Kunitachi,
Tokyo 186-8601, Japan.}
\email{matsushita.y@r.hit-u.ac.jp}
\author{Taisuke Otsu}
\address{Department of Economics, London School of Economics, Houghton Street,
London, WC2A 2AE, UK.}
\email{t.otsu@lse.ac.uk}
\thanks{We are grateful to Naoto Kunitomo for helpful comments.}
\begin{abstract}
This paper extends validity of the conditional likelihood ratio (CLR)
test developed by \textcite{Moreira_2003} to instrumental variable
regression models with unknown error variance and many weak instruments.
In this setting, we argue that the conventional CLR test with estimated
error variance loses exact similarity and is asymptotically invalid.
We propose a modified critical value function for the likelihood ratio
(LR) statistic with estimated error variance, and prove that this
modified test achieves asymptotic validity under many weak instrument
asymptotics. Our critical value function is constructed by representing
the LR using four statistics, instead of two as in \textcite{Moreira_2003}.
A simulation study illustrates the desirable properties of our test.
\end{abstract}

\maketitle

\section{Introduction}

Inference in regression models with endogenous structural variables
and many, weakly relevant instrumental variables is of great importance
in applied research. A canonical example is the paper by \textcite{Angrist_1991},
in which the authors estimate the effect of educational attainment
on wages by constructing up to 1,530 instruments for education through
interacting quarter, state, and year of birth. A more recent literature
that uses many weak instruments employs the ``judge design'' empirical
strategy, which exploits random assignments of cases to judges (e.g.,
\cite{Dahl_2014,Autor_2019}). Since each judge is an instrument in
these settings, and judges can only process a certain number of cases,
the number of instruments typically increases with the sample size.
Many weak instrument settings are also found in papers estimating
the new Keynesian Phillips curves (see \cite{Mavroeidis_2014}) or
studies that employ the Fama-MacBeth method for pricing assets \parencite{Fama_1973}.

In applications involving many weak instruments, researchers often
rely on standard asymptotic approximations when conducting inference.
However, asymptotic approximations to the finite sample distributions
of conventional estimators and test statistics have been shown to
be poor when instruments are weak. The use of many instruments can
improve efficiency of estimators or their associated tests, but often
causes the usual inference procedures to have poor finite sample properties,
and several previous papers have noted this issue. \textcite{Chao_2005},
\textcite{han2006gmm}, and \textcite{newey2009generalized} generalize
the many-instrument asymptotic theory to allow for weak instruments
or moments. \textcite{Andrews_2007} show that the Anderson-Rubin
(AR), Lagrange multiplier (LM), and conditional likelihood ratio (CLR)
tests are robust to many weak instruments, as long as the number of
instruments $k$ grows slower than the cube root of the sample size,
$n^{1/3}$. For the case where $k$ may be proportional to $n$, \textcite{hansen2008estimation}
develop a many-instrument robust standard error and a modification
for the LM test, while \textcite{hausman2012instrumental} propose
Wald tests with limited information maximum likelihood and Fuller
estimators that are robust to heteroskedasticity and many instruments.
More recent developments in conducting robust inference with many
weak instruments include the jackknife AR tests by \textcite{Crudu_2021}
and \textcite{Mikusheva_2021}, and the jackknife LM test by \textcite{matsushita2022jackknife}.

For the weak (but fixed number of) instruments problem, the seminal
work of \textcite{Moreira_2003} sparked a growing literature on conditional
inference. \textcite{Moreira_2003} introduces a general conditional
inference framework for instrumental variable regression models with
homoskedastic errors and advocates the CLR test. \textcite{Andrews_2006}
establish a nearly optimal property of the CLR test, while \textcite{Mills_2014}
propose approximately unbiased conditional Wald tests with comparable
power to the CLR test. \textcite{moreira2019optimal} extend the conditional
inference framework to heteroskedastic and autocorrelated errors.

In this paper, we set out to investigate the finite sample performance
of \citeauthor{Moreira_2003}'s (2003) CLR approach when $k$ is relatively
large, and may grow proportionally with $n$. Size robustness of the
CLR test under $k=o(n^{1/3})$ has already been established by \textcite{Andrews_2007}.
However, we show that in a setting with homoskedastic normal errors
with unknown variance, the conventional CLR test loses exact similarity
and is asymptotically invalid under many weak instrument asymptotics,
where $k$ may grow much faster than $n^{1/3}$. We propose a modified
version of \citeauthor{Moreira_2003}'s (2003) CLR test, hereafter
called the modified CLR (MCLR) test, which is robust to: (i) many
instruments, where the number of instruments can grow at the same
rate or slower than the sample size and (ii) weak instruments, to
the extent that a consistent test may exist \parencite{Mikusheva_2021}.
It should be noted that we use the same test statistic as \textcite{Moreira_2003}
(``$LR_{1}$'' in his paper), but our proposed test employs a different
critical value function which is constructed by representing the likelihood
ratio using four statistics, instead of two as in \textcite{Moreira_2003}.
Our MCLR test retains asymptotic validity when there are many weak
instruments, under the least restrictive condition possible on identification
strength \parencite{Mikusheva_2021}. This result holds even when
we relax the assumption of normally distributed error terms, as long
as we impose an additional moment condition.

The rest of this article is organized as follows. Section \ref{sec:setup}
introduces our setup and the likelihood ratio statistic (LR) when
error variance is unknown, and discusses its representation by four
statistics as well as the properties of those statistics. In Section
\ref{sec:CLR}, we propose our MCLR test by constructing a robust
critical value function and establish its asymptotic validity in a
many weak instruments setting. We also discuss the lack of validity
of the conventional CLR critical value function in our setup. Section
\ref{sec:sim} illustrates the usefulness of our proposed method by
a simulation study. All proofs are contained in Appendix \ref{app:math}.

\section{Setup and test statistics\label{sec:setup}}

\subsection{Setup\label{sub:setup}}

Consider the following instrumental variable regression model:
\begin{eqnarray}
y_{1} & = & Y_{2}\beta+u,\label{eq:model}\\
Y_{2} & = & Z\Pi_{2}+V_{2},\nonumber 
\end{eqnarray}
where $y_{1}=(y_{1i},\ldots,y_{1n})^{\prime}$ is an $n\times1$ vector
of dependent variables, $Y_{2}$ is an $n\times l$ matrix of endogenous
regressors, $\beta$ is an $l\times1$ vector of unknown structural
parameters, $u$ is an $n\times1$ vector of mean-zero disturbances,
$Z$ is an $n\times k$ matrix of instruments, $\Pi_{2}$ is a $k\times l$
matrix of unknown parameters, and $V_{2}$ is an $n\times l$ matrix
of mean-zero error terms. We assume without loss of generality that
there are no exogenous regressors in (\ref{eq:model}) since one can
always partial them out using standard projection methods. The reduced
form system can be written as

\begin{equation}
Y=Z\Pi+V,\label{eq:reduced}
\end{equation}
where $Y=(y_{1},Y_{2})$, $\Pi=(\pi_{1},\Pi_{2})$, and $V=(v_{1},V_{2})$
with $\pi_{1}=\Pi_{2}\beta$ and $v_{1}=V_{2}\beta+u$.

This paper is concerned with testing the null hypothesis $H_{0}:\beta=\beta_{0}$
on the structural parameters, against the alternative $H_{1}:\beta\neq\beta_{0}$,
where the coefficients $\Pi_{2}$ are treated as nuisance parameters.
In particular, this paper focuses on the situation wherein researchers
wish to test such a hypothesis, but only have many weak instruments
at their disposal. 

To proceed, we impose the following assumptions.

\begin{asm}$\quad$ 
\begin{description}
\item [{1}] {[}Normal errors{]} The rows of $V$ are independent and identically
distributed, and follow $N(0,\Omega)$ with a positive definite matrix
$\Omega$. $\Omega$ is unknown to the researcher.
\item [{2.}] {[}Many weak instruments{]} $Z$ is non-random. The number
of instruments $k=k_{n}$ may grow as $n$ increases. One of the following
two conditions hold\\
(a) $\frac{k}{n}\to\alpha\in[0,1)$ as $n\rightarrow\infty$, and
the concentration parameter 
\[
\mu^{2}=(A_{0}^{\prime}\Omega^{-1}A_{0})^{-1/2}A_{0}^{\prime}\Omega^{-1}\Pi^{\prime}Z^{\prime}Z\Pi\Omega^{-1}A_{0}(A_{0}^{\prime}\Omega^{-1}A_{0})^{-1/2},
\]
with $A_{0}=(\beta_{0},I_{l})^{\prime}$ satisfies $\frac{\mu^{2}}{\sqrt{k}}\to\infty$
as $n\rightarrow\infty$; or\\
(b) $\frac{k}{n}\to0$ as $n\rightarrow\infty$ (without any condition
on $\mu^{2}$).
\end{description}
\end{asm}

The normality of reduced-form errors in Assumption 1 is useful to
motivate our conditional inference approach, which is inspired by
the exact similarity of the LR statistic with known $\Omega$. Indeed,
\textcite{Moreira_2003} proves that conditional on the sufficient
statistic $\Pi_{2}$ and when errors are normally distributed, the
LR statistic with known $\Omega$ has a finite-sample distribution
independent of nuisance parameters under $H_{0}$ and its quantiles
can be used to construct a similar test (as long as this distribution
is continuous). Since we maintain \textcite{Moreira_2003}'s conditional
inference framework, we begin with normally distributed error terms,
although we will show that this assumption can be relaxed for the
asymptotic analysis (see Theorem \ref{thm:3}). Throughout this paper,
we focus on the case where $\Omega$ is unknown to researchers. Assumption
2 concerns the instrumental variables. In this paper, we restrict
$Z$ to be non-random, which is equivalent to conditioning on $Z$.
In order to allow for $k$ to grow proportionally with $n$ as in
Assumption 2 (a), we need to impose an additional condition $\frac{\mu^{2}}{\sqrt{k}}\to\infty$,
which imposes a lower bound on the strength of the instruments. Indeed,
$\frac{\mu^{2}}{\sqrt{k}}\to\infty$ is the least restrictive characterization
of weak identification that allows for a consistent test of $H_{0}:\beta=\beta_{0}$
\parencite{Mikusheva_2021}. \textcite{Chao_2005} also impose this
condition to achieve consistency of point estimators under many weak
instrument asymptotics. If $k$ grows slower than $n$, as in Assumption
2 (b), there is no requirement on $\mu^{2}$ , i.e., the instruments
can be arbitrarily weak.

Note that Wald tests based on many-instrument robust standard errors
\parencite{hansen2008estimation,hausman2012instrumental} are asymptotically
valid under Assumption 2 (a), but not under Assumption 2 (b). Our
MCLR test is asymptotically valid in both cases. Simulation studies
in Section \ref{sec:sim} illustrate this distinction numerically.
\textcite{Andrews_2007} show that the conventional CLR test is asymptotically
valid under relatively small numbers of instruments, that is when
$k^{3}/n\to0$. Assumption 2 allows the number of instruments $k$
to be much larger, and as illustrated in our simulation study, the
MCLR test is preferred when $k/n$ is large.

\subsection{Likelihood ratio statistic with known $\Omega$}

We first introduce some notation. When the variance $\Omega$ of $V$
is known, the LR statistic for testing $H_{0}$ against $H_{1}$ is
written as
\begin{equation}
LR_{0}=\frac{b_{0}^{\prime}Y^{\prime}P_{Z}Yb_{0}}{b_{0}^{\prime}\Omega b_{0}}-\bar{\lambda},\label{eq:LR0a}
\end{equation}
where $b_{0}=(1,-\beta_{0}^{\prime})^{\prime}$, $P_{Z}=Z(Z^{\prime}Z)^{-1}Z^{\prime}$
is the projection matrix with respect to $Z$, and $\bar{\lambda}$
is the smallest eigenvalue of $\Omega^{-1/2}Y^{\prime}P_{Z}Y\Omega^{-1/2}$
\parencite{Moreira_2003}.

To derive a more convenient expression for $LR_{0}$, note that $Z^{\prime}Y$
is a sufficient statistic for the parameters $(\beta,\Pi)$ under
the assumption $V\sim N(0,\Omega)$ with known $\Omega$. This implies
that $Z^{\prime}YD$ is also a sufficient statistic, for any nonsingular
matrix $D$. So, we set $D=(b_{0},\Omega^{-1}A_{0})$ and obtain the
partition $Z^{\prime}YD=[S:T]$, where 
\[
S=Z^{\prime}Yb_{0},\qquad T=Z^{\prime}Y\Omega^{-1}A_{0}.
\]
This is a convenient partitioning because $S$ and $T$ are independent
and only $T$ depends on the nuisance parameters $\Pi$. Indeed, under
the null hypothesis, $T$ alone is a sufficient statistic for $\Pi$.

By using standardized versions of $S$ and $T$: 
\[
\bar{S}=(Z^{\prime}Z)^{-1/2}Z^{\prime}Yb_{0}(b_{0}^{\prime}\Omega b_{0})^{-1/2},\qquad\bar{T}=(Z^{\prime}Z)^{-1/2}Z^{\prime}Y\Omega^{-1}A_{0}(A_{0}^{\prime}\Omega^{-1}A_{0})^{-1/2},
\]
the LR statistic $LR_{0}$ can be alternatively written as 
\begin{equation}
LR_{0}=\bar{S}^{\prime}\bar{S}-\bar{\lambda}\equiv\psi_{0}(\bar{S}^{\prime}\bar{S},\bar{S}^{\prime}\bar{T},\bar{T}^{\prime}\bar{T}),\label{eq:LR0b}
\end{equation}
where $\bar{\lambda}$ is the smallest eigenvalue of $(\bar{S},\bar{T})^{\prime}(\bar{S},\bar{T})$.
See Proposition 1 in \textcite{Moreira_2003} for a proof. If $\Omega$
is known, we can apply the conventional CLR test by \textcite{Moreira_2003}
based on $LR_{0}$, even with many weak instruments. Since this paper
focuses on the case of unknown $\Omega$ as stated in Assumption 1,
the conventional test is infeasible. Its feasible counterpart, obtained
by plugging in a consistent estimator of $\Omega$, turns out to be
invalid under the many weak instrument asymptotics (see, Remark \ref{rem:lack}
below).

\subsection{Likelihood ratio statistic with unknown $\Omega$}

We now introduce our test statistic of interest, for the case of unknown
$\Omega$. The error variance matrix $\Omega$ can be estimated by
\begin{equation}
\hat{\Omega}=\frac{1}{n-k}Y^{\prime}M_{Z}Y,\label{eq:Omega}
\end{equation}
where $M_{Z}=I_{n}-P_{Z}$. By replacing $\Omega$ in (\ref{eq:LR0a})
with the estimator $\hat{\Omega}$, the LR statistic for testing $H_{0}$
with unknown $\Omega$ is written as 
\begin{equation}
\frac{LR_{1}}{n-k}=\frac{b_{0}^{\prime}Y^{\prime}P_{Z}Yb_{0}}{b_{0}^{\prime}Y^{\prime}M_{Z}Yb_{0}}-\hat{\lambda},\label{eq:LR1a}
\end{equation}
where $\hat{\lambda}$ is the smallest eigenvalue of $\frac{1}{n-k}\hat{\Omega}^{-1/2}Y^{\prime}P_{Z}Y\hat{\Omega}^{-1/2}$.

To obtain an analogous expression to (\ref{eq:LR0b}) for $LR_{1}$,
we introduce two more objects:
\[
\tilde{S}=M_{Z}Yb_{0}(b_{0}^{\prime}\Omega b_{0})^{-1/2},\qquad\tilde{T}=M_{Z}Y\Omega^{-1}A_{0}(A_{0}^{\prime}\Omega^{-1}A_{0})^{-1/2}.
\]
Based on this notation, we obtain the following representation of
the $LR_{1}$ statistic. 

\begin{prop}\label{prop:1} $LR_{1}$ can be written as a function
of $n$, $k$, and $(\bar{S}^{\prime}\bar{S},\bar{S}^{\prime}\bar{T},\bar{T}^{\prime}\bar{T},\tilde{S}^{\prime}\tilde{S},\tilde{S}^{\prime}\tilde{T},\tilde{T}^{\prime}\tilde{T})$:
\[
\frac{LR_{1}}{n-k}=\psi_{1,n,k}(\bar{S}^{\prime}\bar{S},\bar{S}^{\prime}\bar{T},\bar{T}^{\prime}\bar{T},\tilde{S}^{\prime}\tilde{S},\tilde{S}^{\prime}\tilde{T},\tilde{T}^{\prime}\tilde{T}).
\]
\end{prop}

This proposition says that the LR statistic $LR_{1}$ depends on six
objects, instead of three as for $LR_{0}=\psi_{0}(\bar{S}^{\prime}\bar{S},\bar{S}^{\prime}\bar{T},\bar{T}^{\prime}\bar{T})$
in (\ref{eq:LR0a}). In order to develop our conditional inference
method based on $LR_{1}$, we first establish the following properties
of those six objects.

\begin{prop}\label{prop:2} Under Assumption 1 and the null hypothesis
$H_{0}:\beta=\beta_{0}$, it holds
\begin{description}
\item [{(i)}] $\bar{S}|\bar{T}=t\sim N(0,I_{k})$ and $\bar{S}^{\prime}\bar{T}|\bar{T}=t\sim N(0,t^{\prime}t)$,
\item [{(ii)}] $\left.\left(\begin{array}{cc}
\tilde{S}^{\prime}\tilde{S} & \tilde{S}^{\prime}\tilde{T}\\
\tilde{T}^{\prime}\tilde{S} & \tilde{T}^{\prime}\tilde{T}
\end{array}\right)\right|\bar{T}=t\sim Wishart(n-k,I_{l+1},0)$, 
\item [{(iii)}] $\bar{S}$, $\bar{T}$, and $(\tilde{S},\tilde{T})$ are
mutually independent.
\end{description}
\end{prop}

\begin{rem}\label{rem:suffstat} \textcite{Moreira_2003} builds
a conditional inference framework for the conventional CLR test based
on two sufficient statistics, $\bar{S}$ and $\bar{T}$. We add two
more statistics, $\tilde{S}$ and $\tilde{T}$, which are shown to
be mutually independent of $\bar{S}$ and $\bar{T}$. We need to formally
establish the properties of $\tilde{S}$ and $\tilde{T}$ because
we explicitly focus on the case of unknown $\Omega$, as stated in
Assumption 1. On the other hand, \textcite{Moreira_2003} defines
the conventional CLR test using $LR_{0}$ and later establishes that
using a plug-in consistent estimator for $\Omega$ is asymptotically
valid under weak (but a fixed number of) instruments. Since this will
not be the case under Assumption 2, we directly consider $LR_{1}$.
Under many weak instrument asymptotics, the dimensions of all four
of our statistics $\bar{S}$, $\bar{T}$, $\tilde{S}$ and $\tilde{T}$
grow to $\infty$, which explains why the six objects we focus on
are expressed by inner-products. As we will see in Section \ref{sec:CLR},
$\bar{T}$ will play the most important role in our conditional inference
approach, since it is a sufficient statistic for $\Pi$. Moreover,
$\bar{T}^{\prime}\bar{T}$ is centered at the concentration parameter
$\mu^{2}$, and therefore is a measure of how strongly identified
the first-stage is. We will use this fact in Section \ref{sec:sim}.
\end{rem}

\section{Conditional likelihood ratio test with many weak instruments\label{sec:CLR}}

Based on the test statistic $LR_{1}$ and its properties presented
in the last section, we now develop our conditional inference method.
To begin with, recall that $\bar{T}$ is a sufficient statistic for
$\Pi$, and consider the critical value function for given $\bar{T}=t$:

\[
c_{1,\alpha}(t)=(1-\alpha)\text{-th quantile of }\psi_{1,n,k}(\mathcal{S}^{\prime}\mathcal{S},\mathcal{S}^{\prime}t,t^{\prime}t,\mathcal{W}_{1},\mathcal{W}_{2},\mathcal{W}_{3}),
\]
where $\psi_{1,n,k}$ is defined in Proposition \ref{prop:1}, and
$\mathcal{S}\sim N(0,I_{k})$ and $\left(\begin{array}{cc}
\mathcal{W}_{1} & \mathcal{W}_{2}\\
\mathcal{W}_{2} & \mathcal{W}_{3}
\end{array}\right)\sim Wishart(n-k,I_{l+1},0)$ are independent. Propositions \ref{prop:1} and \ref{prop:2} directly
imply the following property of $c_{1,\alpha}(t)$.

\begin{thm}\label{thm:1} Under Assumption 1 and the null hypothesis
$H_{0}:\beta=\beta_{0}$, it holds that

\begin{equation}
\Pr\left\{ \frac{LR_{1}}{n-k}\ge c_{1,\alpha}(\bar{T})\right\} =1-\alpha.\label{eq:testinfeasible}
\end{equation}
\end{thm}

This theorem says that if $\bar{T}$ is observable, the LR test using
$\frac{LR_{1}}{n-k}$ with the critical value $c_{1,\alpha}(\bar{T})$
is exactly similar. However, since $\bar{T}$ is unobservable for
the case of unknown $\Omega$, the test based on (\ref{eq:testinfeasible})
is infeasible. To develop a feasible version, we estimate $\bar{T}$
by 
\[
\hat{T}=(Z^{\prime}Z)^{-1/2}Z^{\prime}Y\hat{\Omega}^{-1}A_{0}(A_{0}^{\prime}\hat{\Omega}^{-1}A_{0})^{-1/2},
\]
where $\hat{\Omega}$ is as defined in (\ref{eq:Omega}). Based on
this estimator, our proposed rejection rule is defined as: 
\begin{equation}
\text{Reject }H_{0}\text{ if }\frac{LR_{1}}{n-k}\ge c_{1,\alpha}(\hat{T}).\label{eq:test}
\end{equation}
The next theorem is the main result of our paper, and it establishes
asymptotic validity of the MCLR test in (\ref{eq:test}).

\begin{thm} \label{thm:2} Consider the setup in Section \ref{sub:setup}.
Under Assumptions 1 and 2, it holds that
\begin{equation}
\Pr\left\{ \frac{LR_{1}}{n-k}\ge c_{1,\alpha}(\hat{T})\right\} \to1-\alpha\quad\text{as }n\to\infty.\label{eq:consistent}
\end{equation}
\end{thm}

This theorem is derived under the normality assumption (Assumption
1). For non-normal errors, as long as $k/n\to0$, we can also establish
asymptotic validity of the MCLR test by requiring an additional moment
condition. Let $P_{ii}$ be the $(i,i)$-th element of $P_{Z}$.

\begin{thm} \label{thm:3} Consider the setup in Section \ref{sub:setup}.
Under Assumption 2 (b) and $\frac{1}{k}\sum_{i=1}^{n}P_{ii}^{2}\to0$,
(\ref{eq:consistent}) holds true.\end{thm}

Specifically, as long as the number of instruments $k$ grows slower
than the sample size $n$, and $\frac{1}{k}\sum_{i=1}^{n}P_{ii}^{2}\to0$,
our MCLR test is asymptotically valid even when reduced form errors
are non-normal. Note, Assumption 2 (b) is still more general than
$k=o(n^{1/3})$, which \textcite{Andrews_2007} require for asymptotic
validity of the CLR test with non-normal errors.

\begin{rem}\label{rem:lack} {[}Lack of similarity and validity of
conventional CLR test{]} When $\Omega$ is known, the critical value
function of the test statistic $LR_{0}$ in (\ref{eq:LR0b}) for testing
$H_{0}:\beta=\beta_{0}$ can be obtained as 
\[
c_{0,\alpha}(t)=(1-\alpha)\text{-th quantile of }\psi_{0}(\mathcal{D}_{1},\mathcal{D}_{2},t^{\prime}t),
\]
where $\mathcal{D}_{1}\sim\chi^{2}(k-l)$ and $\mathcal{D}_{2}\sim N(0,t^{\prime}t)$
are independent. As shown by \textcite{Moreira_2003}, the test $\mathbb{I}\{LR_{0}\ge c_{0,\alpha}(\bar{T})\}$
is exactly similar for the case of known $\Omega$ (i.e., $\Pr\{LR_{0}\ge c_{0,\alpha}(\bar{T})\}=1-\alpha$).
When $\Omega$ is unknown, \textcite{Moreira_2003} suggested to plug-in
the estimator $\hat{\Omega}$ to the test statistic $LR_{0}$ (which
yields $LR_{1}$) and use $c_{0,\alpha}(\hat{T})$, that is: 
\begin{equation}
\text{Reject }H_{0}\text{ if }LR_{1}\ge c_{0,\alpha}(\hat{T}).\label{eq:conv-test}
\end{equation}
However, since $LR_{1}$ is evidently different from $LR_{0}$, we
cannot guarantee similarity for $LR_{1}$, i.e., 
\[
\Pr\{LR_{1}\ge c_{0,\alpha}(\bar{T})\}\neq\Pr\{LR_{0}\ge c_{0,\alpha}(\bar{T})\}=1-\alpha.
\]
Therefore, even if we ignore the estimation error arising from using
$\hat{T}$ instead of $\bar{T}$, the conventional CLR test in (\ref{eq:conv-test})
is asymptotically invalid in our setup. \end{rem}

\section{Numerical illustrations\label{sec:sim}}

In this section, we compare the critical value function $c_{1,\alpha}(t)$
of the MCLR test with $c_{0,\alpha}(t)$ of the conventional CLR test
(Section \ref{sub:crit}), and use Monte Carlo simulations to evaluate
the finite sample performance of our MCLR test relative to existing
competitors.

\subsection{Critical value function\label{sub:crit}}

The critical value function $c_{1,\alpha}(t)$ of our MCLR test does
not have a closed form, as is the case with \citeauthor{Moreira_2003}'s
(2003) critical value function $c_{0,\alpha}(t)$. Panel A of Table
\ref{tab:crit} presents the critical value function of the MCLR test
for the 5\% significance level. Critical values are shown for $n=100$,
using different values of $\tau=\bar{T}^{\prime}\bar{T}$ and calculated
using 10,000 Monte Carlo replications. Although $c_{1,\alpha}(t)$
takes $\bar{T}$ as input, we choose to focus on $\tau$ for simulations,
which is indicative of identification strength. This also aids comparison
with \textcite{Moreira_2003}, whose critical value function only
depends on $\tau$. 

\begin{figure}[H]
\label{fig:crit}

\caption{Critical value function of MCLR test with $k=4$}

\centering{}\includegraphics[scale=0.6]{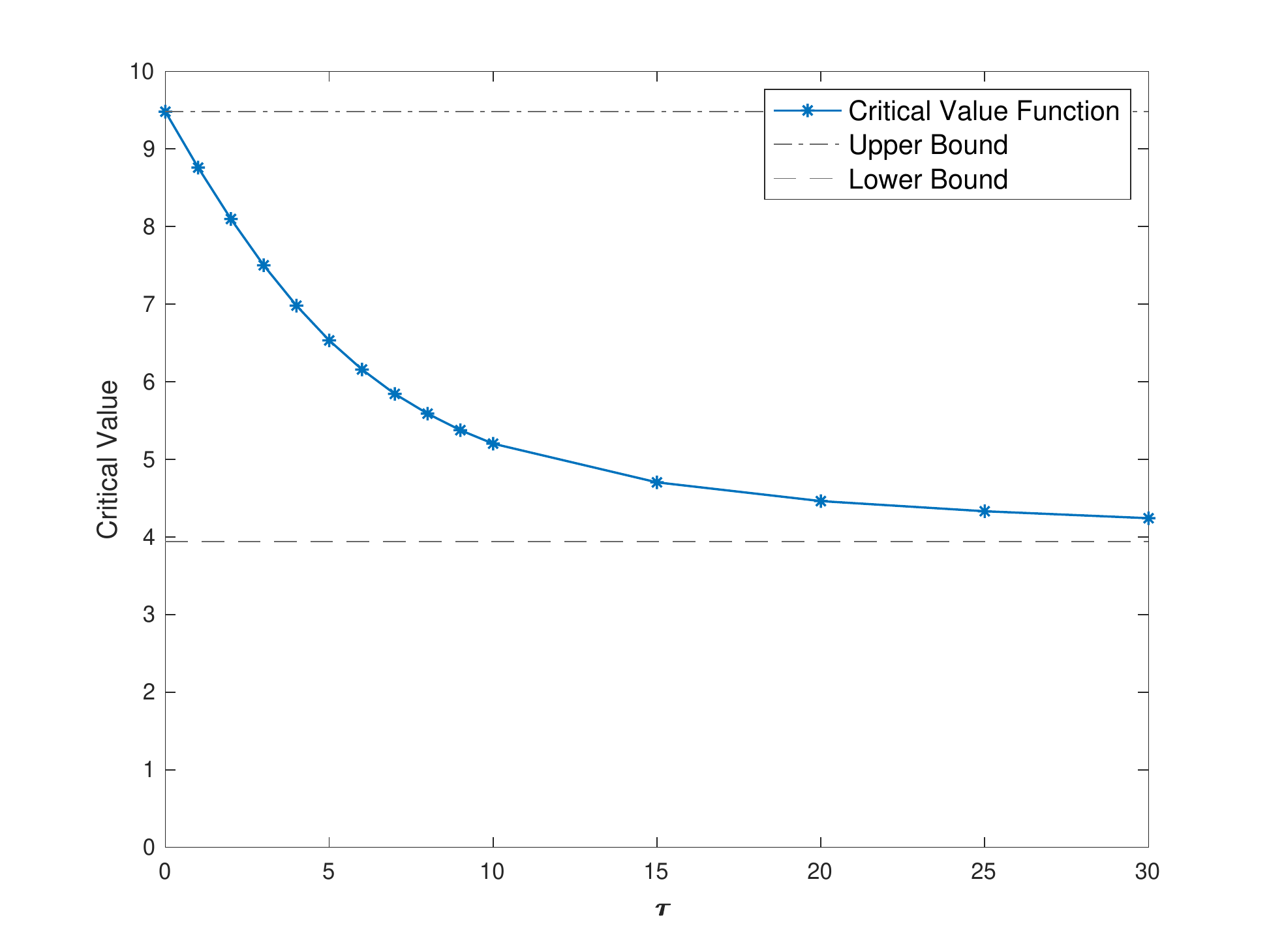} 
\end{figure}

When $k=1$, the critical value function of the MCLR test is a constant
equal to 3.93 for all values of $\tau$, with the slight variation
in the final row being due to simulation error. Just as is the case
for the CLR test, the critical value function of the MCLR test for
any given $k$ has an approximately exponential shape. Figure \ref{fig:crit}
illustrates this with a plot of the critical value function of our
MCLR test when $k=4$. When instruments are weak (i.e., $\tau$ is
small), the critical values are larger. When $\tau$ is larger, such
that instruments are stronger, the test behaves as if it were unconditional
and the critical values are relatively stable around 3.93.

For comparison, in Panel B of Table \ref{tab:crit}, we present the
critical value function $c_{0,\alpha}(t)$ of the conventional CLR
test. As shown in Theorem \ref{thm:2}, this test runs into size problems
when there are many weak instruments. This is evident in the critical
value function - once the number of instruments exceeds a tenth of
the sample size, the critical values of the CLR test lie everywhere
below those of our MCLR test. This suggests that the conventional
CLR test would lead to over-rejection of the null hypothesis $H_{0}:\beta=\beta_{0}$
when the number of instruments is large.

\begin{table}[htb!]
\caption{Critical value functions}
\label{tab:crit}

\begin{tabular}{c|cccccccc}
\multicolumn{9}{c}{Panel A: MCLR}\tabularnewline
\hline 
$\tau$ & $k=1$ & $k=2$ & $k=3$ & $k=4$ & $k=5$ & $k=10$ & $k=20$ & $k=50$\tabularnewline
\hline 
\hline 
1  & 3.93  & 5.72  & 7.46  & 9.13  & 10.75  & 18.45  & 33.09 & 78.94 \tabularnewline
5  & 3.93  & 4.72  & 5.71  & 6.86  & 8.12  & 15.02  & 29.30 & 74.91 \tabularnewline
10  & 3.93  & 4.34  & 4.85  & 5.46  & 6.19  & 11.40  & 24.79 & 70.00\tabularnewline
20  & 3.93  & 4.14  & 4.37  & 4.63  & 4.93  & 7.20  & 16.87 & 60.48\tabularnewline
50  & 3.93  & 4.02  & 4.11  & 4.20  & 4.30  & 4.91  & 7.02 & 35.25\tabularnewline
75  & 3.93  & 3.99  & 4.05  & 4.11  & 4.18  & 4.55  & 5.66 & 20.18\tabularnewline
100  & 3.93  & 3.98  & 4.02  & 4.06  & 4.10  & 4.38  & 5.14 & 12.84\tabularnewline
50000  & 3.94  & 3.94  & 3.94  & 3.94  & 4.10  & 3.94  & 3.96 & 4.04\tabularnewline
\multicolumn{1}{c}{} &  &  &  &  &  &  &  & \tabularnewline
\multicolumn{9}{c}{Panel B: CLR}\tabularnewline
\hline 
$\tau$ & $k=1$ & $k=2$ & $k=3$ & $k=4$ & $k=5$ & $k=10$ & $k=20$ & $k=50$\tabularnewline
\hline 
\hline 
1  & 3.84  & 5.54  & 7.18  & 8.76  & 10.29  & 17.41  & 30.46 & 66.51\tabularnewline
5  & 3.84  & 4.57  & 5.48  & 6.53  & 7.68  & 14.00  & 26.70 & 62.59\tabularnewline
10  & 3.84  & 4.22  & 4.67  & 5.20  & 5.85  & 10.40  & 22.17 & 57.73\tabularnewline
20  & 3.84  & 4.02  & 4.23  & 4.46  & 4.71  & 6.51  & 14.18 & 48.10\tabularnewline
50  & 3.84  & 3.91  & 3.99  & 4.08  & 4.16  & 4.65  & 6.05 & 21.62\tabularnewline
75  & 3.84  & 3.89  & 3.94  & 4.00  & 4.05  & 4.35  & 5.10 & 10.27\tabularnewline
100  & 3.84  & 3.88  & 3.92  & 3.95  & 3.99  & 4.21  & 4.72 & 7.35\tabularnewline
50000  & 3.84  & 3.84  & 3.84  & 3.84  & 3.99  & 3.84  & 3.84 & 3.84\tabularnewline
\hline 
\end{tabular}
\end{table}

\subsection{Simulation}

This subsection conducts a simulation study based on Design I of \textcite{Staiger_1997}
with $l=1$, and $\beta_{0}=0$. $Z$ comprises of: a constant $Z_{1}$,
$Z_{2}$ drawn from $N(0,1)$, $Z_{2}^{2}$, $Z_{2}^{3}$ and $k-4$
elements independently drawn from $N(0,1)$. Once drawn, $Z$ is then
held fixed across simulations. To vary the strength of instruments,
we use a population version of \citeauthor{Stock_2005}'s (2005) pre-test
for weak instruments. We use three different values of $\Pi_{2}$
so that the population first-stage F-statistic $\delta^{2}=\Pi_{2}^{\prime}Z^{\prime}Z\Pi_{2}/\omega_{22}$
takes the values 2 (very weak instruments), 10 (weak instruments),
and 30 (strong instruments), for different values of $k$. The rows
of $(u,V_{2})$ are i.i.d. normal random vectors with unit variances
and correlation $\rho$, which measures the degree of endogeneity
of $Y_{2}$ in (\ref{eq:model}). The number of Monte Carlo replications
is 10,000 for the size analysis and 2,500 for the power analysis.

Table \ref{tab:sim1-size} investigates the size properties of five
tests for $n=100$ and $H_{0}:\beta=\beta_{0}$: (i) the t-test with
the heteroskedasticity robust limited information maximum likelihood
estimator by \textcite{hausman2012instrumental} (H-LIML), (ii) the
conditional likelihood ratio test by \textcite{Moreira_2003} (CLR),
(iii) the modified Lagrange multiplier test by \textcite{hansen2008estimation}
(mKLM), (iv) the jackknifed version of the Anderson--Rubin (AR) test
proposed by \textcite{Mikusheva_2021} (J-AR) and our proposed modified
CLR test (MCLR). Firstly, note that the size distortions of H-LIML
are large, except when $\delta^{2}$ is large relative to $k$. The
degree of endogeneity of $Y_{2}$ also seems to matter; when $\rho=0.2$,
the t-test tends to under-reject the null hypothesis, while when $\rho=0.6$,
the null is over-rejected. The distortions of the test are most severe
when $\delta^{2}$ is small relative to $k$, and $k$ is large. As
a result, we do not investigate power of this test.

The CLR test attains roughly the correct size when $k=5$, even when
identification is weak and the degree of endogeneity is high. However,
size distortions can be observed when $k/n\ge0.1$. Surprisingly,
this is not visibly exacerbated when $\delta^{2}$ is low, suggesting
that it is the existence of many instruments that has the more severe
empirical consequences on the conventional CLR test. Overall, even
when the CLR test experiences little size distortion, it always has
empirical rejection frequency farther from $5\%$ than our proposed
MCLR test. 

\begin{table}[htb!]
\begin{centering}
\label{tab:sim1-size}
\par\end{centering}
\caption{Empirical rejection frequencies at 5\% significance level}

\begin{tabular}{lcc|ccccc}
\hline 
\multicolumn{1}{c}{\textbf{$\rho$}} & \multicolumn{1}{c}{\textbf{$\delta^{2}$}} & \multicolumn{1}{c|}{\textbf{$k$}} & \multicolumn{1}{c}{H-LIML} & \multicolumn{1}{c}{CLR} & \multicolumn{1}{c}{mKLM} & \multicolumn{1}{c}{J-AR} & \multicolumn{1}{c}{M-CLR}\tabularnewline
\hline 
\hline 
0.2  & 30  & 5  & 0.038  & 0.055  & 0.051 & 0.063 & 0.048\tabularnewline
0.2  & 30  & 10  & 0.030 & 0.064 & 0.050 & 0.064 & 0.054\tabularnewline
0.2  & 30  & 30  & 0.028  & 0.092 & 0.048  & 0.064 & 0.055\tabularnewline
\hline 
0.2  & 10  & 5  & 0.022  & 0.046 & 0.046 & 0.060 & 0.042\tabularnewline
0.2  & 10  & 10  & 0.018 & 0.059 & 0.045 & 0.056 & 0.047\tabularnewline
0.2  & 10  & 30  & 0.016  & 0.093 & 0.044 & 0.060 & 0.048\tabularnewline
\hline 
0.2  & 2  & 5  & 0.008  & 0.048  & 0.041 & 0.062 & 0.042\tabularnewline
0.2  & 2  & 10  & 0.007  & 0.057 & 0.039 & 0.052 & 0.039\tabularnewline
0.2  & 2  & 30  & 0.013 & 0.087 & 0.046  & 0.058 & 0.050\tabularnewline
\hline 
0.6  & 30  & 5  & 0.058  & 0.053 & 0.055  & 0.062 & 0.048\tabularnewline
0.6  & 30  & 10  & 0.058 & 0.062 & 0.053 & 0.061 & 0.052\tabularnewline
0.6  & 30  & 30  & 0.052 & 0.076  & 0.048  & 0.056 & 0.048\tabularnewline
\hline 
0.6  & 10  & 5  & 0.070  & 0.049 & 0.047  & 0.062 & 0.045\tabularnewline
0.6  & 10  & 10  & 0.069 & 0.062 & 0.050 & 0.056 & 0.056\tabularnewline
0.6  & 10  & 30  & 0.086 & 0.092 & 0.047 & 0.057 & 0.048\tabularnewline
\hline 
0.6  & 2  & 5  & 0.084 & 0.050  & 0.042  & 0.065 & 0.046\tabularnewline
0.6  & 2  & 10  & 0.092 & 0.058 & 0.040  & 0.059 & 0.037\tabularnewline
0.6  & 2  & 30  & 0.106 & 0.088 & 0.045  & 0.060 & 0.057\tabularnewline
\hline 
\end{tabular}
\end{table}

The mKLM test works well. Although it tends to under-reject when $\delta^{2}$
is relatively small, its size distortions never exceed $1\%$. Similarly,
the J-AR test appears relatively robust to many weak instruments,
although it tends to over-reject the null in all cases. 

Compared to the other tests that we consider, the rejection frequencies
of our proposed MCLR test are on average closest to the nominal level.
As our theory in Section \ref{sec:CLR} suggests, the MCLR test is
robust to weak instruments, many instruments, and many weak instruments. 

\begin{figure}[H]
\label{fig:1}\caption{Calibrated Power Curves}

\textcolor{white}{hello}

\includegraphics[scale=0.44]{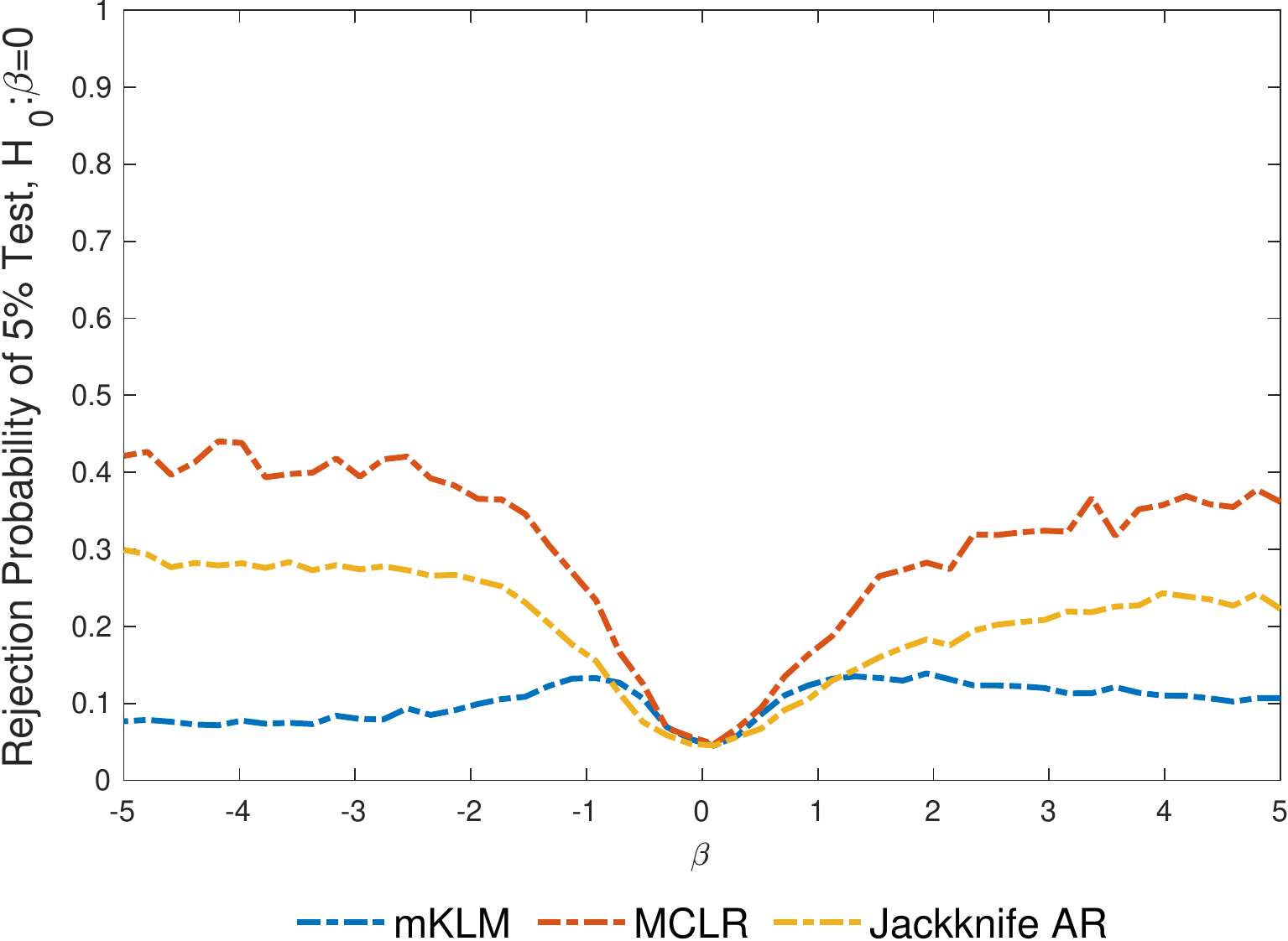}$\qquad$\includegraphics[scale=0.4]{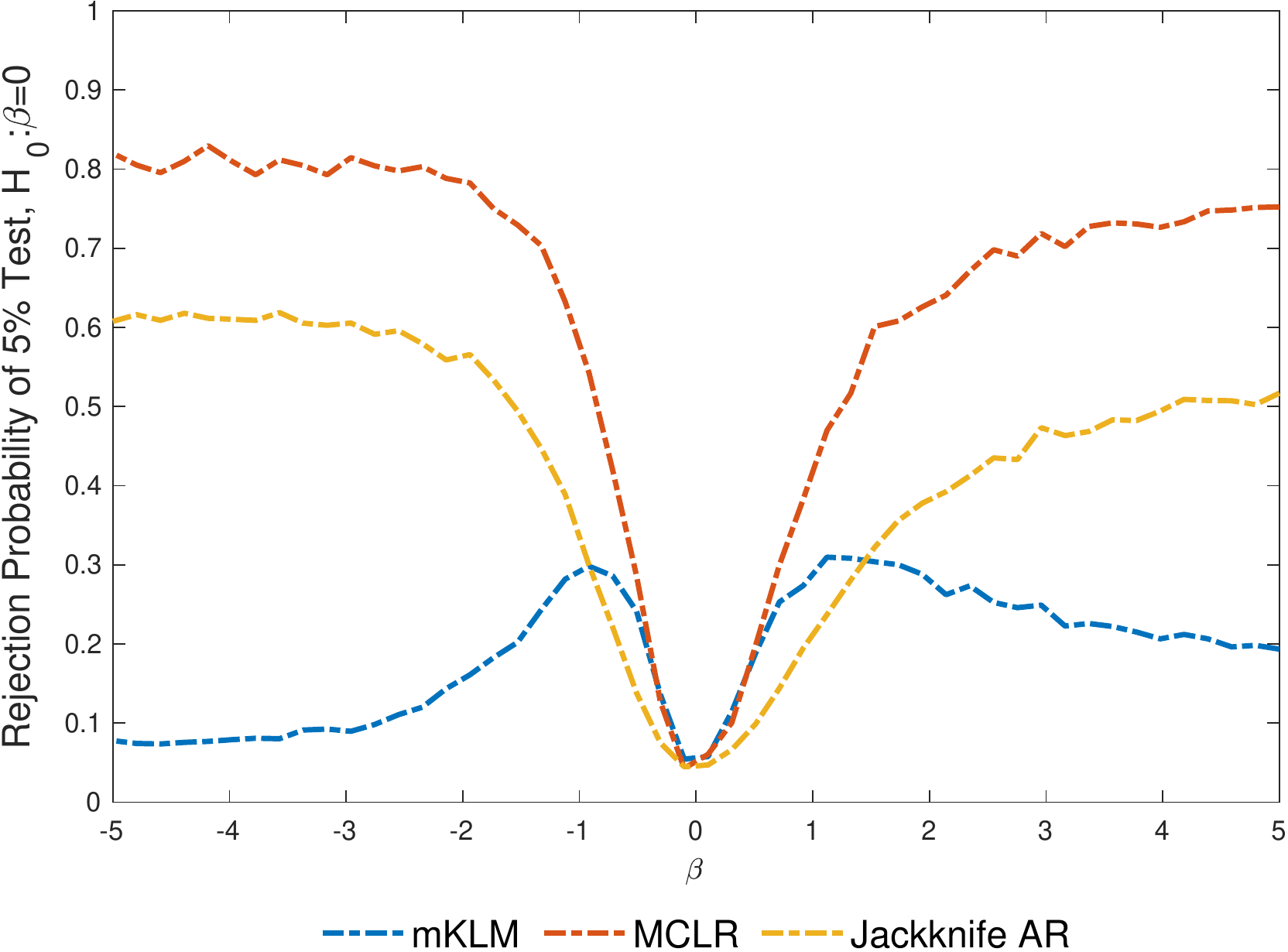}\\
\includegraphics[scale=0.4]{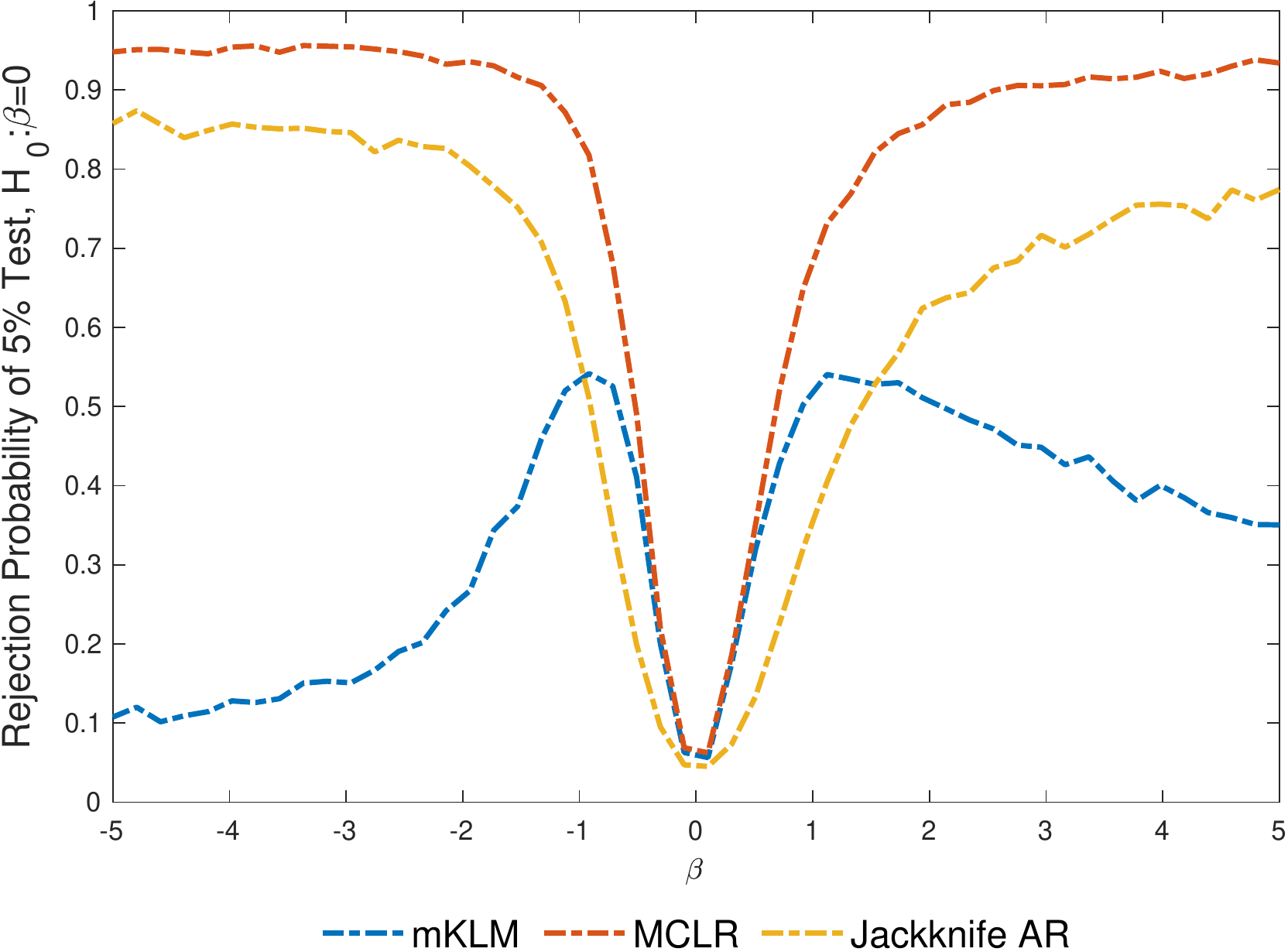}$\qquad$\includegraphics[scale=0.4]{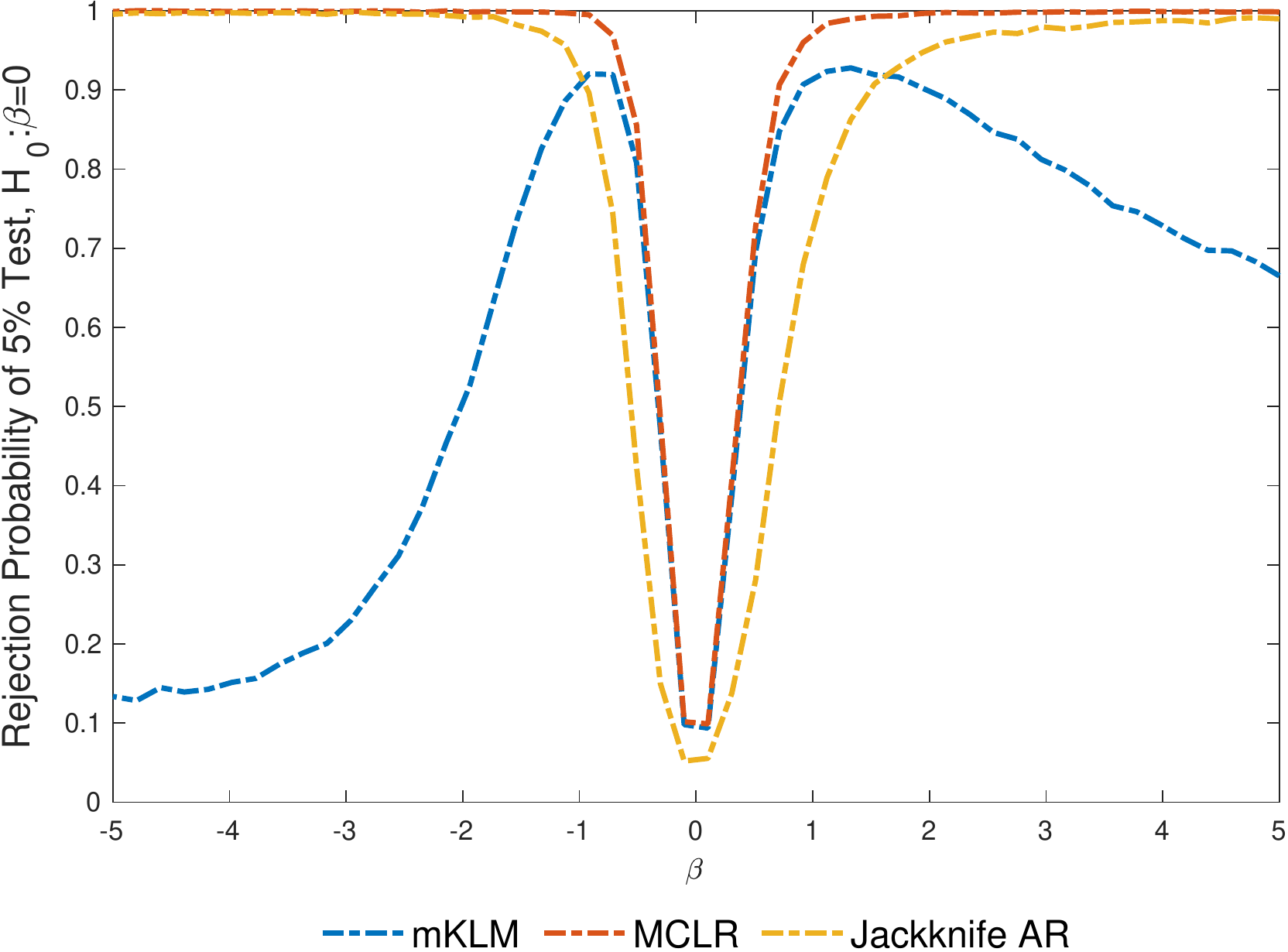}\\
Note: From left to right, and top to bottom, these figures plot power
curves for: $\delta^{2}/k=1/3$, $\delta^{2}/k=1/2$, $\delta^{2}/k=1$
and $\delta^{2}/k=2$, with $\rho=0.2$ and $n=200$. 
\end{figure}

Figure \ref{fig:1} presents calibrated power curves for the MCLR,
J-AR and mKLM tests for $H_{0}:\beta=\beta_{0}$, under the alternative
hypotheses $H_{1}:\beta=\beta_{0}+\Delta$. These power curves are
plotted with respect to the $5\%$ significance level, i.e., the critical
values for these three tests are given by the $95$-th percentiles
of these test statistics under \textbf{$H_{0}$}, computed via 10,000
Monte Carlo replications. Each curve is plotted for $n=200$, $l=1$,
and $\beta_{0}=0$. We present four different cases, with different
values of $\delta^{2}/k$ , holding $\rho=0.2$. As we move from left
to right, and top to bottom, the figures show the cases of $\delta^{2}/k=1/3$,
$1/2$, $1$, and $2$. Our MCLR test is uniformly more powerful at
all values of $\delta^{2}/k$, although is more pronounced for low
$\delta^{2}/k$. The mKLM test experiences spurious declines in power
under alternative hypotheses that are further away from the null,
with consistently low power when $\delta^{2}/k=1/3$. When $\delta^{2}/k$
is high, i.e., identification is strong and/or the number of instruments
is small, the J-AR has similar, although everywhere lower power, than
our M-CLR test. While we do not present theoretical results on power,
this result suggests that the MCLR test shares the superior power
properties of the conventional CLR test, which has near optimal power
with small $k$ \parencite{Andrews_2006}. 

\section{Conclusion\label{sec:conclusion}}

In this paper, we propose a modification of \citeauthor{Moreira_2003}'s
(2003) conditional likelihood ratio (CLR) test, namely the MCLR test.
We prove that in instrumental variable regression models with unknown
error variance and many weak instruments, the MCLR test is asymptotically
valid under many weak instrument asymptotics, unlike the CLR test.
This is true even when the number of instruments grows proportionally
to the sample size, and identification is as weak as possible to the
extent that a consistent test may exist \parencite{Mikusheva_2021}. 

Our simulation study indicates that the MCLR test has superior size
properties to the CLR test and is more powerful than competing tests
that are robust to many weak instruments, including the modified Lagrange
multiplier test by \textcite{hansen2008estimation} and the jackknife
Anderson--Rubin test by \textcite{Mikusheva_2021}.

The size and power results presented in our theorems and simulation
study lead us to recommend the MCLR test for general use in scenarios
when instruments are many and potentially weak. This is based on the
fact that the MCLR test is asymptotically valid with little size distortion
in finite samples with many weak instruments, while also retaining
the favorable power properties of the CLR test.

\newpage{}

\appendix

\section{Mathematical appendix\label{app:math}}

\subsection{Proof of Proposition \ref{prop:1}}

Let $(D_{1},\ldots,D_{6})=(\bar{S}^{\prime}\bar{S},\bar{S}^{\prime}\bar{T},\bar{T}^{\prime}\bar{T},\tilde{S}^{\prime}\tilde{S},\tilde{S}^{\prime}\tilde{T},\tilde{T}^{\prime}\tilde{T})$.
Recall that $\frac{LR_{1}}{n-k}=\frac{b_{0}^{\prime}Y^{\prime}P_{Z}Yb_{0}}{b_{0}^{\prime}Y^{\prime}M_{Z}Yb_{0}}-\hat{\lambda}$,
where $\hat{\lambda}$ is the smallest eigenvalue of $(n-k)^{-1}\hat{\Omega}^{-1/2}Y^{\prime}P_{Z}Y\hat{\Omega}^{-1/2}$.
The numerator of the first term can be written as 
\[
b_{0}^{\prime}Y^{\prime}P_{Z}Yb_{0}=(b_{0}^{\prime}\Omega b_{0})\bar{S}^{\prime}\bar{S}=(b_{0}^{\prime}\Omega b_{0})D_{1},
\]
where the first equality follows from the definition of $\bar{S}$.
Similarly, the denominator of the first term of $\frac{LR_{1}}{n-k}$
can be written as 
\[
b_{0}^{\prime}Y^{\prime}M_{Z}Yb_{0}=(b_{0}^{\prime}\Omega b_{0})\tilde{S}^{\prime}\tilde{S}=(b_{0}^{\prime}\Omega b_{0})D_{4},
\]
where the first equality follows from the definition of $\tilde{S}$.
Thus the first term of $\frac{LR_{1}}{n-k}$ is written as $\frac{D_{1}}{D_{4}}$.

We now consider the second term of $\frac{LR_{1}}{n-k}$. Observe
that $\hat{\lambda}$ is the minimum eigenvalue solution of $|\hat{\Omega}^{-1/2}Y^{\prime}P_{Z}Y\hat{\Omega}^{-1/2}-\hat{\lambda}I|=0$,
or equivalently 
\[
\left|F^{\prime}Y^{\prime}P_{Z}YF-\frac{\hat{\lambda}}{n-k}F^{\prime}Y^{\prime}M_{Z}YF\right|=0,
\]
for any nonsingular matrix $F$. By setting $F=[b_{0}(b_{0}^{\prime}\Omega b_{0})^{-1/2}:\Omega^{-1}A_{0}(A_{0}^{\prime}\Omega^{-1}A_{0})^{-1/2}]$,
the above equation can be written as 
\[
0=\left|\left(\begin{array}{cc}
\bar{S}^{\prime}\bar{S} & \bar{S}^{\prime}\bar{T}\\
\bar{T}^{\prime}\bar{S} & \bar{T}^{\prime}\bar{T}
\end{array}\right)-\frac{\hat{\lambda}}{n-k}\left(\begin{array}{cc}
\tilde{S}^{\prime}\tilde{S} & \tilde{S}^{\prime}\tilde{T}\\
\tilde{T}^{\prime}\tilde{S} & \tilde{T}^{\prime}\tilde{T}
\end{array}\right)\right|=\left|\left(\begin{array}{cc}
D_{1} & D_{2}\\
D_{2}^{\prime} & D_{3}
\end{array}\right)-\frac{\hat{\lambda}}{n-k}\left(\begin{array}{cc}
D_{4} & D_{5}\\
D_{5}^{\prime} & D_{6}
\end{array}\right)\right|.
\]
Therefore, $\hat{\lambda}$ can be solved for as a function of $(D_{1},\ldots,D_{6})$.
Combining these results, we obtain the conclusion.

\subsection{Proof of Proposition \ref{prop:2}}

\subsubsection{Proof of (i)}

As shown in \textcite{Moreira_2003}, $\bar{S}\sim N(0,I_{K})$ and
$\bar{S}$ and $\bar{T}$ are independent. Therefore, the conclusion
follows.

\subsubsection{Proof of (ii)}

Observe that 
\[
\left(\begin{array}{cc}
\tilde{S}^{\prime}\tilde{S} & \tilde{S}^{\prime}\tilde{T}\\
\tilde{T}^{\prime}\tilde{S} & \tilde{T}^{\prime}\tilde{T}
\end{array}\right)=W^{\prime}M_{Z}W,
\]
where $W=[Vb_{0}(b_{0}^{\prime}\Omega b_{0})^{-1/2}:V\Omega^{-1}A_{0}(A_{0}^{\prime}\Omega^{-1}A_{0})^{-1/2}]$.
Since $M_{Z}$ is an $n\times n$ non-random idempotent matrix with
$\mathrm{rank}(M_{Z})=n-k$, it is sufficient for the conclusion to
show that given $\bar{T}=t$, the rows of $W$ are i.i.d. $N(0,I_{l+1})$.
The $i$-th row of $W$ can be written as 
\begin{equation}
W_{i}^{\prime}=[V_{i}^{\prime}b_{0}(b_{0}^{\prime}\Omega b_{0})^{-1/2}:V_{i}^{\prime}\Omega^{-1}A_{0}(A_{0}^{\prime}\Omega^{-1}A_{0})^{-1/2}],\label{pf:0}
\end{equation}
where the $i$-th row $V_{i}^{\prime}$ of $V$ satisfies $V_{i}\sim N(0,\Omega)$.
Thus, we can see that $Var(V_{i}^{\prime}b_{0}(b_{0}^{\prime}\Omega b_{0})^{-1/2})=1$,
$Var(V_{i}^{\prime}\Omega^{-1}A_{0}(A_{0}^{\prime}\Omega^{-1}A_{0})^{-1/2})=I_{l}$,
and $Cov(V_{i}^{\prime}b_{0}(b_{0}^{\prime}\Omega b_{0})^{-1/2},V_{i}^{\prime}\Omega^{-1}A_{0}(A_{0}^{\prime}\Omega^{-1}A_{0})^{-1/2})=0$,
and the conclusion follows.

\subsubsection{Proof of (iii)}

Since $\bar{S}$ and $\bar{T}$ are independent, it is sufficient
to show that $(\bar{S},\bar{T})$ and $(\tilde{S},\tilde{T})$ are
independent. Note that $[\tilde{S}:\tilde{T}]=M_{Z}W$ whose $i$-th
row $W_{i}$ is defined in (\ref{pf:0}). Since 
\[
\bar{S}=(b_{0}^{\prime}\Omega b_{0})^{-1/2})^{-1/2}(Z^{\prime}Z)^{-1/2}Z^{\prime}(v_{1}-V_{2}\beta_{0}),
\]
we have $Cov(\bar{S},[\tilde{S}:\tilde{T}])=0$. Also since both $\bar{S}$
and $[\tilde{S}:\tilde{T}]$ are normal, we obtain independence of
$(\bar{S},\bar{T})$ and $(\tilde{S},\tilde{T})$.

\subsection{Proof of Theorem \ref{thm:2}}

To simplify the presentation, we provide the proof for the case of
$l=1$. Analogous arguments hold for $l>1$. Let 
\[
\mathcal{S}\sim N(0,I_{k}),\qquad\left(\begin{array}{cc}
\mathcal{W}_{1} & \mathcal{W}_{2}\\
\mathcal{W}_{2} & \mathcal{W}_{3}
\end{array}\right)\sim Wishart(n-k,I_{2},0),
\]
be drawn independently, and define 
\[
\Psi(t)=\psi_{1,n,k}(\mathcal{S}^{\prime}\mathcal{S},\mathcal{S}^{\prime}t,t^{\prime}t,\mathcal{W}_{1},\mathcal{W}_{2},\mathcal{W}_{3}),
\]
so that the critical value function for given $\bar{T}=t$ is given
by $c_{1,\alpha}(t)$, the $(1-\alpha)$-th quantile of $\Psi(t)$.

\subsubsection{Proof under Assumption 2 (a)}

For the conclusion in (\ref{eq:consistent}), it is sufficient to
show that 
\begin{align}
 & (n-k)\frac{\mu^{2}}{k}\Psi(\bar{T})\text{ converges to some non-degenerate distribution},\label{pf:1}\\
 & (n-k)\frac{\mu^{2}}{k}\{\Psi(\hat{T})-\Psi(\bar{T})\}\overset{p}{\to}0.\label{pf:2}
\end{align}

We first show (\ref{pf:1}). By explicitly computing the smallest
eigenvalue in $\Psi(t)$, $\Psi(\bar{T})$ can be written as 
\begin{equation}
\Psi(\bar{T})=\frac{\mathcal{S}^{\prime}\mathcal{S}}{\mathcal{W}_{1}}+\frac{b+\sqrt{b^{2}-4ac}}{2a},\label{pf:3}
\end{equation}
where 
\begin{eqnarray*}
a & = & \frac{1}{(n-k)^{2}}(\mathcal{W}_{1}\mathcal{W}_{3}-\mathcal{W}_{2}^{2}),\\
b & = & \frac{1}{(n-k)^{2}}\{-\mathcal{W}_{1}(\bar{T}^{\prime}\bar{T})-(\mathcal{S}^{\prime}\mathcal{S})\mathcal{W}_{3}+2(\mathcal{S}^{\prime}\bar{T})\mathcal{W}_{2}\},\\
c & = & \frac{1}{(n-k)^{2}}\{(\mathcal{S}^{\prime}\mathcal{S})(\bar{T}^{\prime}\bar{T})-(\mathcal{S}^{\prime}\bar{T})^{2}\}.
\end{eqnarray*}
To proceed, we express $a$, $b$, and $c$ by the following standardized
objects 
\begin{eqnarray*}
\mathcal{Z}_{1} & = & \frac{\mathcal{S}^{\prime}\mathcal{S}-k}{\sqrt{k}},\qquad\mathcal{Z}_{2}=\frac{1}{\sqrt{k}}\mathcal{S}^{\prime}\bar{T},\qquad\mathcal{Z}_{\bar{T}}=\frac{\bar{T}^{\prime}\bar{T}-k-\mu^{2}}{\sqrt{k}},\\
\mathcal{Q}_{1} & = & \frac{\mathcal{W}_{1}-(n-k)}{\sqrt{n-k}},\qquad\mathcal{Q}_{2}=\frac{\mathcal{W}_{2}}{\sqrt{n-k}},\qquad\mathcal{Q}_{3}=\frac{\mathcal{W}_{3}-(n-k)}{\sqrt{n-k}},
\end{eqnarray*}
where $\mu^{2}=(A_{0}^{\prime}\Omega^{-1}A_{0})^{-1/2}A_{0}^{\prime}\Omega^{-1}\Pi^{\prime}Z^{\prime}Z\Pi\Omega^{-1}A_{0}(A_{0}^{\prime}\Omega^{-1}A_{0})^{-1/2}$
is the concentration parameter defined in Assumption 2 (a). Based
on this notation, $a$, $b$, and $c$ are written as
\begin{eqnarray*}
a & = & 1+\frac{\mathcal{Q}_{1}+\mathcal{Q}_{3}}{\sqrt{n-k}}+\frac{\mathcal{Q}_{1}\mathcal{Q}_{3}-\mathcal{Q}_{2}^{2}}{n-k},\\
b & = & -\frac{k}{n-k}\left\{ 2+\frac{\mu^{2}}{k}+\frac{\left(1+\frac{\mu^{2}}{k}\right)\mathcal{Q}_{1}+\mathcal{Q}_{3}}{\sqrt{n-k}}+\frac{\mathcal{Z}_{1}+\mathcal{Z}_{\bar{T}}}{\sqrt{k}}+\frac{\mathcal{Z}_{\bar{T}}\mathcal{Q}_{1}+\mathcal{Z}_{1}\mathcal{Q}_{3}-2\mathcal{Z}_{2}\mathcal{Q}_{2}}{\sqrt{k}\sqrt{n-k}}\right\} ,\\
c & = & \frac{k^{2}}{(n-k)^{2}}\left\{ 1+\frac{\mu^{2}}{k}+\frac{\left(1+\frac{\mu^{2}}{k}\right)\mathcal{Z}_{1}+\mathcal{Z}_{\bar{T}}}{\sqrt{k}}+\frac{\mathcal{Z}_{1}\mathcal{Z}_{\bar{T}}-\mathcal{Z}_{2}^{2}}{k}\right\} .
\end{eqnarray*}
Based on these expressions and by using $\frac{\mu^{2}}{\sqrt{k}}\to\infty$
(Assumption 2 (a)), we can expand the second term of $\Psi(\bar{T})$
in (\ref{pf:3}), that is 
\begin{eqnarray}
 &  & \frac{-b-\sqrt{b^{2}-4ac}}{2a}\nonumber \\
 & = & \left[(A_{0}+A_{1}+A_{2})-B_{0}\left\{ 1+\frac{1}{2}(B_{1}+B_{2})-\frac{1}{8}B_{1}^{2}\right\} \right]\{1-(C_{1}+C_{2})+C_{1}^{2}\}+o_{p}(n^{-1})\nonumber \\
 & = & (A_{0}-B_{0})+\left\{ A_{1}-\frac{1}{2}B_{0}B_{1}-(A_{0}-B_{0})C_{1}\right\} \nonumber \\
 &  & +\left\{ A_{2}-\frac{1}{2}B_{0}B_{2}+\frac{1}{8}B_{0}B_{1}^{2}+(A_{0}-B_{0})(-C_{2}+C_{1}^{2})-(A_{1}-\frac{1}{2}B_{0}B_{1})C_{1}\right\} +o_{p}(n^{-1}),\label{pf:4}
\end{eqnarray}
where 
\begin{eqnarray*}
A_{0} & = & \frac{1}{2}\frac{k}{n-k}\left(2+\frac{\mu^{2}}{k}\right),\qquad A_{1}=\frac{1}{2}\frac{k}{n-k}\left\{ \frac{\mathcal{Z}_{1}+\mathcal{Z}_{\bar{T}}}{\sqrt{k}}+\frac{\left(1+\frac{\mu^{2}}{k}\right)\mathcal{Q}_{1}+\mathcal{Q}_{3}}{\sqrt{n-k}}\right\} ,\\
A_{2} & = & \frac{1}{2}\frac{k}{n-k}\left(\frac{\mathcal{Z}_{\bar{T}}\mathcal{Q}_{1}+\mathcal{Z}_{1}\mathcal{Q}_{3}-2\mathcal{Z}_{2}\mathcal{Q}_{2}}{\sqrt{k}\sqrt{n-k}}\right),\\
B_{0} & = & \frac{1}{2}\frac{k}{n-k}\frac{\mu^{2}}{k},\qquad B_{1}=\left(\frac{\mu^{2}}{k}\right)^{-1}\left\{ -\frac{2\mathcal{Z}_{1}}{\sqrt{k}}+\frac{2\mathcal{Z}_{\bar{T}}}{\sqrt{k}}+\frac{2\left(1+\frac{\mu^{2}}{k}\right)\mathcal{Q}_{1}}{\sqrt{n-k}}-\frac{2\mathcal{Q}_{3}}{\sqrt{n-k}}\right\} ,\\
B_{2} & = & \left(\frac{\mu^{2}}{k}\right)^{-2}\left[\begin{array}{c}
\frac{\left(1+\frac{\mu^{2}}{k}\right)^{2}}{n-k}\mathcal{Q}_{1}^{2}+\frac{4+4\frac{\mu^{2}}{k}}{n-k}\mathcal{Q}_{2}^{2}+\frac{1}{n-k}\mathcal{Q}_{3}^{2}-\frac{2+2\frac{\mu^{2}}{k}}{n-k}\mathcal{Q}_{1}\mathcal{Q}_{3}\\
+\frac{1}{k}\mathcal{Z}_{\bar{T}}^{2}+\frac{1}{k}\mathcal{Z}_{1}^{2}+\frac{4}{k}\mathcal{Z}_{2}^{2}-\frac{2}{k}\mathcal{Z}_{1}\mathcal{Z}_{\bar{T}}\\
+\frac{2+4\frac{\mu^{2}}{k}}{\sqrt{k}\sqrt{n-k}}\mathcal{Z}_{\bar{T}}\mathcal{Q}_{1}-\frac{2}{\sqrt{k}\sqrt{n-k}}\mathcal{Z}_{\bar{T}}\mathcal{Q}_{3}\\
-\frac{2+2\frac{\mu^{2}}{k}}{\sqrt{k}\sqrt{n-k}}\mathcal{Z}_{1}\mathcal{Q}_{1}+\frac{2-2\frac{\mu^{2}}{k}}{\sqrt{k}\sqrt{n-k}}\mathcal{Z}_{1}\mathcal{Q}_{3}-\frac{8+4\frac{\mu^{2}}{k}}{\sqrt{k}\sqrt{n-k}}\mathcal{Z}_{2}\mathcal{Q}_{2}
\end{array}\right],\\
C_{1} & = & \frac{\mathcal{Q}_{1}+\mathcal{Q}_{3}}{\sqrt{n-k}},\qquad C_{2}=\frac{\mathcal{Q}_{1}\mathcal{Q}_{3}-\mathcal{Q}_{2}^{2}}{n-k}.
\end{eqnarray*}
Also an expansion of the first term of (\ref{pf:3}) is

\[
\frac{\mathcal{S}^{\prime}\mathcal{S}}{\mathcal{W}_{1}}=\frac{k}{n-k}\left\{ 1+\frac{\mathcal{Z}_{1}}{\sqrt{k}}\right\} \left\{ 1-\frac{\mathcal{Q}_{1}}{\sqrt{n-k}}+\frac{\mathcal{Q}_{1}^{2}}{n-k}\right\} +o_{p}(n^{-1}).
\]
Combining these results with direct calculations under Assumption
2 (a), an expansion for $\Psi(\bar{T})$ is obtained as
\begin{eqnarray}
(n-k)\frac{\mu^{2}}{k}\Psi(\bar{T}) & = & \frac{k}{n-k}\mathcal{Q}_{2}^{2}+\frac{\sqrt{k}}{n-k}\frac{\mu^{2}}{k}\mathcal{Z}_{1}\mathcal{Q}_{1}^{2}-2\sqrt{\frac{k}{n-k}}\mathcal{Z}_{2}\mathcal{Q}_{2}+\mathcal{Z}_{2}^{2}+o_{p}(1)\nonumber \\
 & = & \left(\mathcal{Z}_{2}-\sqrt{\frac{\alpha}{1-\alpha}}\mathcal{Q}_{2}\right)^{2}+o_{p}(1),\label{pf:5}
\end{eqnarray}
where the second equality follows from $\frac{k}{n}\to\alpha$ and
$\frac{\mu^{2}}{k}=O(1)$. Since $(\mathcal{Z}_{2},\mathcal{Q}_{2})$
converges to a non-degenerate distribution, we obtain (\ref{pf:1}).

We now show (\ref{pf:2}). A similar argument to derive (\ref{pf:5})
(by replacing $\bar{T}$ with $\hat{T}$) yields that
\[
(n-k)\frac{\mu^{2}}{k}\Psi(\hat{T})=\left(\hat{\mathcal{Z}}_{2}-\sqrt{\frac{\alpha}{1-\alpha}}\mathcal{Q}_{2}\right)^{2}+o_{p}(1),
\]
where $\hat{\mathcal{Z}}_{2}=\frac{1}{\sqrt{k}}\mathcal{S}^{\prime}\hat{T}$.
Thus it is sufficient for (\ref{pf:1}) to show that 
\begin{equation}
\hat{\mathcal{Z}}_{2}=\mathcal{Z}_{2}+o_{p}(1).\label{pf:6}
\end{equation}
Let 
\[
F=[b_{0}(b_{0}^{\prime}\Omega b_{0})^{-1/2}:\Omega^{-1}A_{0}(A_{0}^{\prime}\Omega^{-1}A_{0})^{-1/2}].
\]
Note that 
\begin{eqnarray*}
\mathcal{S}^{\prime}\hat{T} & = & \mathcal{S}^{\prime}(Z^{\prime}Z)^{-1/2}Z^{\prime}Y\hat{\Omega}^{-1}A_{0}(A_{0}^{\prime}\hat{\Omega}^{-1}A_{0})^{-1/2}\\
 & = & \sqrt{n-k}\mathcal{S}^{\prime}(Z^{\prime}Z)^{-1/2}Z^{\prime}Y(Y^{\prime}M_{Z}Y)^{-1}A_{0}(A_{0}^{\prime}(Y^{\prime}M_{Z}Y)^{-1}A_{0})^{-1/2}\\
 & = & \sqrt{n-k}\mathcal{S}^{\prime}(Z^{\prime}Z)^{-1/2}Z^{\prime}YF\left(\begin{array}{cc}
\tilde{S}^{\prime}\tilde{S} & \tilde{S}^{\prime}\tilde{T}\\
\tilde{T}^{\prime}\tilde{S} & \tilde{T}^{\prime}\tilde{T}
\end{array}\right)^{-1}F^{\prime}A_{0}\left(A_{0}^{\prime}F\left(\begin{array}{cc}
\tilde{S}^{\prime}\tilde{S} & \tilde{S}^{\prime}\tilde{T}\\
\tilde{T}^{\prime}\tilde{S} & \tilde{T}^{\prime}\tilde{T}
\end{array}\right)^{-1}F^{\prime}A_{0}\right)^{-1/2}\\
 & = & \sqrt{n-k}(\mathcal{S}^{\prime}\bar{S},\mathcal{S}^{\prime}\bar{T})\left(\begin{array}{cc}
\tilde{S}^{\prime}\tilde{S} & \tilde{S}^{\prime}\tilde{T}\\
\tilde{T}^{\prime}\tilde{S} & \tilde{T}^{\prime}\tilde{T}
\end{array}\right)^{-1}\left(\begin{array}{c}
0\\
\left\{ \frac{1}{(\tilde{S}^{\prime}\tilde{S})(\tilde{T}^{\prime}\tilde{T})-(\tilde{S}^{\prime}\tilde{T})^{2}}(\tilde{S}^{\prime}\tilde{S})\right\} ^{-1/2}
\end{array}\right)\\
 & = & \sqrt{n-k}\left\{ \frac{1}{(\tilde{S}^{\prime}\tilde{S})(\tilde{T}^{\prime}\tilde{T})-(\tilde{S}^{\prime}\tilde{T})^{2}}\right\} ^{1/2}\left\{ -(\mathcal{S}^{\prime}\bar{S})(\tilde{S}^{\prime}\tilde{T})(\tilde{S}^{\prime}\tilde{S})^{-1/2}+(\mathcal{S}^{\prime}\bar{T})(\tilde{S}^{\prime}\tilde{S})^{1/2}\right\} .
\end{eqnarray*}
Thus, we have
\begin{eqnarray*}
\hat{\mathcal{Z}}_{2} & = & \frac{\sqrt{k}}{\sqrt{n-k}}\left[\frac{1}{\left\{ \frac{\tilde{S}^{\prime}\tilde{S}-(n-k)}{n-k}+1\right\} \left\{ \frac{\tilde{T}^{\prime}\tilde{T}-(n-k)}{n-k}+1\right\} -\frac{1}{n-k}\left(\frac{\tilde{S}^{\prime}\tilde{T}}{\sqrt{n-k}}\right)^{2}}\right]^{1/2}\\
 &  & \times\left\{ -\frac{1}{\sqrt{k}}\left(\frac{\mathcal{S}^{\prime}\bar{S}}{\sqrt{k}}\right)\left(\frac{\tilde{S}^{\prime}\tilde{T}}{\sqrt{n-k}}\right)\left(\frac{\tilde{S}^{\prime}\tilde{S}-(n-k)}{n-k}+1\right)^{-1/2}\right.\\
 &  & \left.+\frac{\sqrt{n-k}}{\sqrt{k}}\left(\frac{\mathcal{S}^{\prime}\bar{T}}{\sqrt{k}}\right)\left(\frac{\tilde{S}^{\prime}\tilde{S}-(n-k)}{n-k}+1\right)^{1/2}\right\} \\
 & = & \mathcal{Z}_{2}+o_{p}(1),
\end{eqnarray*}
i.e., (\ref{pf:6}) holds true. Therefore, we obtain (\ref{pf:2}).
Since (\ref{pf:1}) and (\ref{pf:2}) are satisfied, the conclusion
follows.

\subsubsection{Proof under Assumption 2 (b)}

By an analogous argument in the proof of Moreira (2003, Theorem 2),
we can see that the conclusion under Assumption 2 (b) follows by:
\begin{eqnarray}
\hat{\mathcal{Z}}_{2} & = & \mathcal{Z}_{2}+o_{p}(1),\label{pf:1a}\\
\hat{\mathcal{Z}}_{\hat{T}} & = & \mathcal{Z}_{\bar{T}}+o_{p}(1).\label{pf:2a}
\end{eqnarray}
For (\ref{pf:1a}), we can apply the same argument as the proof of
(\ref{pf:6}) (since it does not use the condition on $\mu^{2}$).
For (\ref{pf:2a}), note that
\begin{eqnarray*}
\frac{1}{k}\hat{T}^{\prime}\hat{T} & = & (A_{0}^{\prime}\hat{\Omega}^{-1}A_{0})^{-1/2}A_{0}^{\prime}\hat{\Omega}^{-1}Y^{\prime}P_{Z}Y\hat{\Omega}^{-1}A_{0}(A_{0}^{\prime}\hat{\Omega}^{-1}A_{0})^{-1/2}\\
 & = & (n-k)(A_{0}^{\prime}(Y^{\prime}M_{Z}Y)^{-1}A_{0})^{-1/2}A_{0}^{\prime}(Y^{\prime}M_{Z}Y)^{-1}Y^{\prime}P_{Z}Y(Y^{\prime}M_{Z}Y)^{-1}A_{0}(A_{0}^{\prime}(Y^{\prime}M_{Z}Y)^{-1}A_{0})^{-1/2}\\
 & = & \frac{n-k}{k}\{(\tilde{S}^{\prime}\tilde{S})(\tilde{T}^{\prime}\tilde{T})-(\tilde{S}^{\prime}\tilde{T})^{2}\}^{-1}\left(\begin{array}{c}
-(\tilde{S}^{\prime}\tilde{T})(\tilde{S}^{\prime}\tilde{S})^{-1/2}\\
(\tilde{S}^{\prime}\tilde{S})^{1/2}
\end{array}\right)^{\prime}\left(\begin{array}{cc}
\bar{S}^{\prime}\bar{S} & \bar{S}^{\prime}\bar{T}\\
\bar{S}^{\prime}\bar{T} & \bar{T}^{\prime}\bar{T}
\end{array}\right)\left(\begin{array}{c}
-(\tilde{S}^{\prime}\tilde{T})(\tilde{S}^{\prime}\tilde{S})^{-1/2}\\
(\tilde{S}^{\prime}\tilde{S})^{1/2}
\end{array}\right)\\
 & = & \frac{n-k}{k}\{(\tilde{S}^{\prime}\tilde{S})(\tilde{T}^{\prime}\tilde{T})-(\tilde{S}^{\prime}\tilde{T})^{2}\}^{-1}\left\{ (\bar{S}^{\prime}\bar{S})(\tilde{S}^{\prime}\tilde{T})^{2}(\tilde{S}^{\prime}\tilde{S})^{-1}-2(\bar{S}^{\prime}\bar{T})(\tilde{S}^{\prime}\tilde{T})+(\bar{T}^{\prime}\bar{T})(\tilde{S}^{\prime}\tilde{S})\right\} \\
 & = & \left[\left\{ \frac{\tilde{S}^{\prime}\tilde{S}-(n-k)}{n-k}+1\right\} \left\{ \frac{\tilde{T}^{\prime}\tilde{T}-(n-k)}{n-k}+1\right\} -\frac{1}{n-k}\left(\frac{\tilde{S}^{\prime}\tilde{T}}{\sqrt{n-k}}\right)^{2}\right]^{-1}\\
 &  & \times\left\{ \frac{1}{n-k}\left(\frac{\bar{S}^{\prime}\bar{S}-k}{k}+1\right)\left(\frac{\tilde{S}^{\prime}\tilde{T}}{\sqrt{n-k}}\right)^{2}\left(\frac{\tilde{S}^{\prime}\tilde{S}-(n-k)}{n-k}+1\right)^{-1}\right.\\
 &  & \left.-\frac{2}{\sqrt{k}\sqrt{n-k}}\left(\frac{\bar{S}^{\prime}\bar{T}}{\sqrt{k}}\right)\left(\frac{\tilde{S}^{\prime}\tilde{T}}{\sqrt{n-k}}\right)+\left(\frac{\bar{T}^{\prime}\bar{T}-\mu^{2}-k}{k}+\frac{\mu^{2}}{k}+1\right)\left(\frac{\tilde{S}^{\prime}\tilde{S}-(n-k)}{n-k}+1\right)\right\} \\
 & = & \frac{\bar{T}^{\prime}\bar{T}-\mu^{2}-k}{k}+\left(\frac{\mu^{2}}{k}+1\right)\frac{\tilde{S}^{\prime}\tilde{S}-(n-k)}{n-k}+\left(\frac{\mu^{2}}{k}+1\right)+O_{p}(K^{-1}),
\end{eqnarray*}
where the first five equalities follow from the definitions and direct
algebra, and the last equality follows from $\frac{\tilde{S}^{\prime}\tilde{T}}{\sqrt{n-k}}=O_{p}(1)$,
$\frac{\bar{S}^{\prime}\bar{S}-k}{\sqrt{k}}=O_{p}(1)$, $\frac{\tilde{S}^{\prime}\tilde{S}-(n-k)}{\sqrt{n-k}}=O_{p}(1)$,
and $\frac{\bar{S}^{\prime}\bar{T}}{\sqrt{k}}=O_{p}(1)$. Therefore,
(\ref{pf:2a}) is verified as:
\[
\hat{\mathcal{Z}}_{\hat{T}}=\mathcal{Z}_{\bar{T}}+\sqrt{\frac{k}{n-k}}\left(\frac{\mu^{2}}{k}+1\right)\frac{\tilde{S}^{\prime}\tilde{S}-(n-k)}{\sqrt{n-k}}+o_{p}(1)=\mathcal{Z}_{\bar{T}}+o_{p}(1),
\]
where the second equality follows from $k/n\to0$ (Assumption 2 (b)). 

\subsection{Proof of Theorem \ref{thm:3}}

Under $k/n\to0$ (Assumption 2 (b)), $\left(\frac{\tilde{S}^{\prime}\tilde{S}-(n-k)}{\sqrt{n-k}},\frac{\tilde{S}^{\prime}\tilde{T}}{\sqrt{n-k}},\frac{\tilde{T}^{\prime}\tilde{T}-(n-k)}{\sqrt{n-k}}\right)$
are of smaller order than $(\bar{\mathcal{Z}}_{1},\bar{\mathcal{Z}}_{2},\bar{\mathcal{Z}}_{\bar{T}})=\left(\frac{\bar{S}^{\prime}\bar{S}-k}{\sqrt{k}},\frac{\bar{S}^{\prime}\bar{T}}{\sqrt{k}},\frac{\bar{T}^{\prime}\bar{T}-k-\mu^{2}}{\sqrt{k}}\right)$.
Thus, it is enough for the conclusion to show that non-normality of
the errors does not affect the limit of the variance, i.e., 

\[
Var(\bar{\mathcal{Z}}_{1},\bar{\mathcal{Z}}_{2},\bar{\mathcal{Z}}_{\bar{T}})\to\left(\begin{array}{ccc}
2 & 0 & 0\\
0 & 1 & 0\\
0 & 0 & 2
\end{array}\right).
\]

To see this, let $u_{i}=Y_{i}^{\prime}b_{0}$, $w_{i}=Y_{i}^{\prime}\Omega^{-1}A_{0}$,
$\sigma_{u}^{2}=Var(u_{i})$, $\kappa_{u}=E[u_{i}^{4}]$, and $P_{ij}$
be the $(i,j)$-th element of $P_{Z}$. We have 
\begin{eqnarray*}
Var(\bar{\mathcal{Z}}_{1}) & = & \frac{1}{k\sigma_{u}^{4}}\left\{ E\left[\left(\sum_{i=1}^{n}\sum_{j=1}^{n}u_{i}u_{j}P_{ij}\right)^{2}\right]-\left(E\left[\sum_{i=1}^{n}\sum_{j=1}^{n}u_{i}u_{j}P_{ij}\right]\right)^{2}\right\} \\
 & = & \frac{1}{k\sigma_{u}^{4}}\left\{ E\left[\sum_{i=1}^{n}\sum_{j\neq i}^{n}u_{i}^{2}u_{j}^{2}(2P_{ij}^{2}+P_{ii}P_{jj})+\sum_{i=1}^{n}u_{i}^{4}P_{ii}^{2}\right]-\left(E\left[\sum_{i=1}^{n}u_{i}^{2}P_{ii}\right]\right)^{2}\right\} \\
 & = & \frac{1}{k\sigma_{u}^{4}}\left\{ \sigma_{u}^{4}\left\{ \sum_{i=1}^{n}\sum_{j=1}^{n}(2P_{ij}^{2}+P_{ii}P_{jj})-3\sum_{i=1}^{n}P_{ii}^{2}\right\} +\kappa_{u}\sum_{i=1}^{n}P_{ii}^{2}-\sigma_{u}^{4}\left(\sum_{i=1}^{n}P_{ii}\right)^{2}\right\} \\
 & = & \frac{1}{k\sigma_{u}^{4}}\left\{ 2k\sigma_{u}^{4}+(\kappa_{u}-3\sigma_{u}^{4})\sum_{i=1}^{n}P_{ii}^{2}\right\} \to2,
\end{eqnarray*}
where the fourth equality follows from $\sum_{i=1}^{n}\sum_{j=1}^{n}P_{ij}^{2}=\sum_{i=1}^{n}P_{ii}=k$,
and the convergence follows from the assumption $\frac{1}{k}\sum_{i=1}^{n}P_{ii}^{2}\to0$.
Similarly, letting $\sigma_{w}^{2}=Var(w_{i})$, we have 
\begin{eqnarray*}
Cov(\bar{\mathcal{Z}}_{1},\bar{\mathcal{Z}}_{2}) & = & \frac{1}{k\sigma_{u}^{3}\sigma_{w}}\left\{ E\left[\left(\sum_{i=1}^{n}\sum_{j=1}^{n}u_{i}u_{j}P_{ij}\right)\left(\sum_{i=1}^{n}\sum_{j=1}^{n}u_{i}w_{j}P_{ij}\right)\right]\right.\\
 &  & \left.-E\left[\sum_{i=1}^{n}\sum_{j=1}^{n}u_{i}u_{j}P_{ij}\right]E\left[\sum_{i=1}^{n}\sum_{j=1}^{n}u_{i}w_{j}P_{ij}\right]\right\} \\
 & = & \frac{1}{k\sigma_{u}^{3}\sigma_{w}}E\left[\left(\sum_{i=1}^{n}\sum_{j=1}^{n}u_{i}u_{j}P_{ij}\right)\left(\sum_{i=1}^{n}\sum_{j=1}^{n}u_{i}w_{j}P_{ij}\right)\right]\\
 & = & \frac{1}{k\sigma_{u}^{3}\sigma_{w}}E\left[\sum_{i=1}^{n}u_{i}^{3}w_{i}P_{ii}^{2}\right]=\frac{E[u_{i}^{3}w_{i}]}{\sigma_{u}^{2}\sigma_{uw}}\frac{1}{k}\sum_{i=1}^{n}P_{ii}^{2}\to0,
\end{eqnarray*}
where the second equality follows from $E[u_{i}w_{i}]=0$.

The limits of the other elements can be shown in the same manner,
so we obtain the conclusion.

\printbibliography

\end{document}